\documentclass{vgtc}                          % final (conference style)
\graphicspath{{figures/}{pictures/}{images/}{./}} % where to search for the images

\usepackage{times}                     % we use Times as the main font
\usepackage{tabu}                      % only used for the table example
\usepackage{booktabs}                  % only used for the table example
\usepackage{lipsum}                    % used to generate placeholder text
\usepackage{mwe}                       % used to generate placeholder figures

\usepackage{mathptmx}                  % use matching math font
\usepackage{enumitem}
\onlineid{0}

\vgtccategory{Research}

\vgtcinsertpkg

\title{Characterizing Multimodal Interaction in Visualization Authoring Tools}

\author{Astrid van den Brandt\\%
        \scriptsize a.v.d.brandt@tue.nl\\
        \scriptsize TU Eindhoven%
\and Sehi L'Yi\\%
\scriptsize sehi\_lyi@hms.harvard.edu\\
     \scriptsize Harvard University%
\and Huyen N. Nguyen\\%
\scriptsize huyen\_nguyen@hms.harvard.edu\\
     \scriptsize Harvard University%
     \and \\Anna Vilanova\\%
        \scriptsize \phantom{blablabla}a.vilanova@tue.nl\phantom{blablabla}\\
     \scriptsize TU Eindhoven%
\and \\Nils Gehlenborg\\%
\scriptsize nils@hms.harvard.edu\\
     \scriptsize Harvard University}
\abstract{
    Multimodal interaction has been increasingly considered in designing visualization authoring tools.
    However, multimodal interaction has a broad meaning in visualization authoring, according to our literature review. 
    Although some previous studies compare different authoring tools, a comprehensive overview of the diverse characteristics of multimodal interaction in visualization authoring tools is still missing. This paper seeks to offer a systematic perspective on how multimodal interaction is integrated within visualization authoring tools. Such an overview can enhance understanding of current practices, highlight distinguishing features among tools, and help identify future research directions, guiding designers in developing more accessible and effective authoring systems.
    We review \nrTools visualization authoring tools that incorporate multimodal interaction and characterize how multimodal interaction is applied in these tools.
    Based on the review results, we discuss design implications and future directions.
    The survey results are also accessible via \url{https://gosling-lang.github.io/multimodal-vis-authoring/}.
} % end of abstract

\keywords{Visualization authoring, multimodal interaction, design space}

\usepackage{xspace}

\PassOptionsToPackage{svgnames}{xcolor}
\definecolor{orange}{HTML}{E69F00}
\definecolor{blue}{HTML}{0072B2}
\definecolor{green}{HTML}{009E73}
\definecolor{pink}{HTML}{CC79A7}
\definecolor{red}{HTML}{D55D00}

\definecolor{NavyBlue}{HTML}{010086}

\newcommand{\nrTools}{20\xspace}

\usepackage{svg}
\newcommand{\orcid}[1]{\href{https://orcid.org/#1}{\includesvg[width=10pt]{figures/orcid.logo.icon}}}
\begin{document}


\firstsection{Introduction}

\maketitle

Visualization researchers have explored the use of tools to make data visualization and creation of custom visualizations easier for a diverse audience~\cite{satyanarayan_critical_2019}. These visualization authoring tools---also referred to as visualization construction tools---typically support users in creating \textit{visual structures} and \textit{views} from \textit{data tables}~\cite{card_1999, grammel_survey_2013}~(\cref{fig:card}). Authoring visualizations has been recognized as a non-trivial task, especially for visualization novices or domain experts who sometimes lack formal training in programming and human--computer interaction (HCI)~\cite{ bigelow_reflections_2014, grammel_how_2010}. The challenge to support users with diverse needs and backgrounds has motivated researchers to study and design different types of authoring interfaces, such as template editors, shelf construction tools, and interactive code editors~\cite{grammel_how_2010}. 

Besides interface design, another aspect that is increasingly considered in visualization authoring tools is multimodal interaction. 
There is ongoing research into multimodal interaction for improving human--computer interfaces, inspired by the fact that people naturally interact with the world through multiple perceptual modalities~\cite{turk_multimodal_2014}. Multimodal interaction has been found to decrease errors and increase the flexibility of systems and user satisfaction~\cite{oviatt_ten_1999}. 
For visualization authoring, it is found that multimodal input, such as touch and speech~\cite{srinivasan_orko_2018}, mouse, keyboard, and speech~\cite{setlur_eviza_2016}, or pen and touch~\cite{jo_touchpivot_2017}, offer various benefits, including freedom of expression~\cite{lee_beyond_2012, srinivasan_inchorus_2020} and more natural, fluid interaction~\cite{srinivasan_orko_2018}.
Accessibility research has investigated multimodal interaction to reach broader audiences~\cite{lee_multimodal_2018, zong_umwelt_2024}. Next to input modalities, researchers have been exploring how to support multimodal outputs, such as providing textual or sound descriptions along with visualization, to enable an effective authoring process for people with vision disabilities~\cite{zong_umwelt_2024}. 
Similarly, a few researchers have been exploring ways of combining multiple authoring interfaces, such as shelves and visualization-by-demonstration~\cite{saket_liger_2019, saket_visualization_2017} or template-editing, shelf-construction, and code-editing~\cite{mcnutt_integrated_2021}, to leverage benefits that individual modalities offer in expressive power, efficiency, and usability of systems. Overall, we find evidence that multimodal interaction can enhance various aspects of authoring tools, making them more effective for creating custom visualizations. 

In surveying multimodal examples, we found that supporting multimodal interaction has a broad meaning in visualization authoring, referring to both input modalities (e.g., keyboard input, mouse, touch, and speech) and interface interaction techniques (e.g., template-editing, shelf configuration, code editing, and natural language interpreters) aimed at overcoming challenges related to usability, accessibility, and authoring flexibility. While individual papers draw some comparisons between tools, we currently lack an overview of the multimodal characteristics of visualization authoring tools. 
This paper aims to provide a structured view of how multimodal interaction is applied in visualization authoring tools. This overview can help to gain a better understanding of how multimodal interaction is currently utilized in authoring systems and what characterizes tools relative to others, to identify research opportunities and guide designers in creating tools that are widely accessible~\cite{lee_reaching_2020} and tailored to specific authoring tasks. To achieve this, we systematically reviewed literature describing multimodal interaction for visualization authoring. Based on our review of \nrTools papers, we created a design space of 5 dimensions---input modalities, interface modalities, authoring tasks, cooperation methods and output modalities---to characterize multimodal interaction for visualization authoring. The design space forms the basis to discuss design implications, open questions, and future research directions. 
\section{Methodology}
We systematically reviewed papers using the selection criteria and a paper collection approach. Additionally, we drew upon existing frameworks to further scope and analyze multimodal authoring.

\subsection{Selection criteria}

We focus on data visualization authoring tools, which we defined as systems that support at least changing the visualization type through (a set of) encoding (or visual mapping) tasks~\cite{card_1999}. Using this criterion, we excluded tools for data transformation only (e.g., Wrangler~\cite{kandel_wrangler_2011}) and systems that mainly focus on data exploration with data encoding or reconfiguration options (e.g., Orko~\cite{srinivasan_orko_2018}). Additionally, we excluded visual tools that are not for \emph{data} visualizations, such as creative programming, generative arts, designing, or photo and video editing. 

We included papers describing designs, prototypes and full-fledged implementations of visualization authoring tools with multimodal interaction. Furthermore, we considered tools supporting multimodal interaction if it was claimed in the paper by the authors. Lastly, we excluded commercial tools for the reason that they often lack details about the tasks and modalities supported.

\subsection{Paper collection}
Paper collection was performed with three approaches. (1) We performed keyword search for publications at major Visualization and HCI venues, such as VIS, TVCG, and CHI, using Google Scholar on Apr 30, 2024 with the following keywords: (\texttt{"multimodal interaction" "visualization" "authoring")}. We also performed searches with the commonly used alternative spelling \texttt{"multi-modal interaction"}. Using this keyword search strategy resulted in a collection of 7 seed papers meeting the earlier specified criteria. 
Second, we used a snowballing approach~\cite{claes_snowballing_2014} to collect papers cited by these seed papers. This resulted in 7 papers being added. Finally, we added 6 papers through a search \textit{in the wild}, where we followed the work of leading authors, the literature that builds upon their contributions, and papers that were known to us. The paper collection process yielded \nrTools relevant papers.

\subsection{Existing Classifications}\label{sec:classifications}
We make use of three existing taxonomies and classifications as a starting point to characterize the authoring tasks, interface modalities and cooperation methods in the collected papers, forming the basis for our design space.

\begin{figure}[!t]
    \centering
    \includegraphics[width=1\linewidth, alt={}]{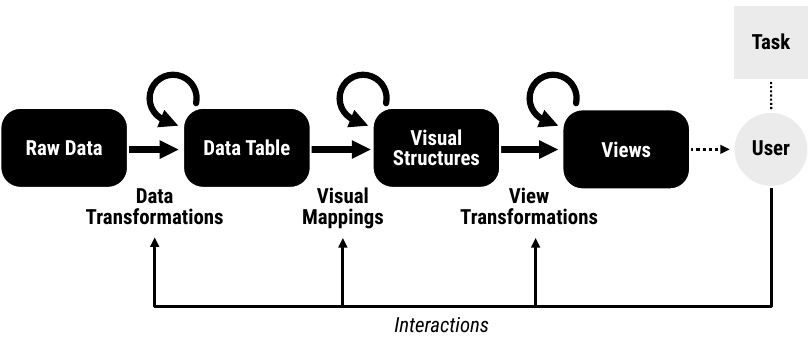}
    \caption{%
    The Visualization Reference Model~\cite{card_1999}. We adopted the model to identify visualization authoring tasks in our survey.
    }
    \label{fig:card}
    % \vspace{-5mm}
\end{figure}

\paragraph{Authoring Tasks.}
Similar to Grammel et al.~\cite{grammel_survey_2013}, we use the reference visualization model by Card et al.~\cite{card_1999} (\cref{fig:card}) to define the types of authoring tasks. The model describes three high-level user tasks during the visualization creation process:
(1) \textbf{Data Transformations}: Users transform raw data into idiosyncratic formats (\textit{data tables)} suitable for visualization. This task includes subtasks such as filtering, aggregations, calculations, and derivations of new data;
(2) \textbf{Visual Mappings}: Users encode the data as \textit{visual structures} during the visual mapping task. This task involves, amongst others, attribute selection, mark and channel selection, and choosing among visualization types;
(3) \textbf{View Transformations}: Users can modify or augment the display of \textit{visual structures} by graphical parameters creating \textit{views} in space-time, i.e., showing \textit{visual structures} from different perspectives or with varying levels of detail. This includes brushing and linking, showing details-on-demand, and other interactions such as zooming, panning and clipping.

\paragraph{Interface Modalities.}
To code the types of interface modalities, we referred to common methods in the landscape of visualization construction tools surveyed by Grammel et al.~\cite{grammel_survey_2013} and additionally included two
interfaces that gained popularity for visualization authoring after the survey’s
publication:
(1) \textbf{Template Editors} allow users to author familiar visualizations using templates that feature a predefined set of chart types (e.g., RAWGraphs~\cite{mauri_rawgraphs_2017}); 
(2) \textbf{Shelf Configuration} interfaces allow more bottom-up authoring process by letting users map data fields to visual channels, often with a drag-and-drop interaction (e.g.,  Polaris~\cite{stolte_polaris_2007}, now Tableau);
(3) \textbf{Natural Language Interfaces} \textbf{(NLI}) enable users to author visualizations through a conversational AI agent (e.g., VisTalk~\cite{wang_towards_2022}); (4) \textbf{Code Editors} (or Textual Programming interfaces) allow users to specify visualizations with code or structured text (e.g., JSON grammars such as Vega-Lite~\cite{satyanarayan_vega-lite_nodate});
% (5) \textbf{Visualization-by-demonstration (VbD)}~\cite{saket_visualization_2017}, a direct manipulation-based (DM)  technique mapping user intent to authoring operations (e.g., DataBreeze~\cite{srinivasan_interweaving_2021})
(5) \textbf{Visualization-by-demonstration (VbD)}~\cite{saket_visualization_2017}, a direct manipulation-based (DM)  technique that allows users to edit visualizations by directly interacting (e.g., using mouse or pen) with the graphical encodings in the visual representation. The system interprets the user's intent from these interactions and maps it authoring operations (e.g., DataBreeze~\cite{srinivasan_interweaving_2021});
% (6) \textbf{Visualization-by-example (VbE)} approaches are related to template editing, but they allow users to submit an example (e.g., an image or textual specification), from which the system generates a reusable visualization template (e.g., AutoGosling~\cite{wang_enabling_2023}); 
and (6) \textbf{Visual Builders} facilitate construction of custom visualizations by letting users select primitives from a palette, assemble and bind them to data in a canvas area, similar to graphics editing software (e.g., Data Illustrator~\cite{dataillustrator});

\paragraph{Cooperation Methods.}
Our categorization in this section is informed by the taxonomy of cooperative multimodal interactions proposed by Marin et al.~\cite{martin_tycoon_nodate}, which distinguishes types such as transfer, equivalence, specialization, redundancy, and complementarity. However, we focus on the subset of cooperation methods most relevant to multimodal authoring, where modalities work together to represent or explore data. We exclude redundancy and transfer as we find these are more relevant for system-level coordination. Furthermore, we explicitly distinguish between two subtypes of complementary methods to better capture the nuances of modality.
We describe four distinct ways in which two or more input and interface modalities can be combined:
(1) \textbf{Equivalence} is used when several modalities act as alternatives to achieve a task, enabling adaptation to user preferences. For example, InChorus~\cite{srinivasan_inchorus_2020} allows binding attributes to encodings with all three modalities (pen, touch and speech);
% (2) \orange{\textbf{Redundancy} cooperation means that a system can handle redundant task specifications by the user often to enable faster interaction. An example is when users select a town both by speech and touch (Siroux et al. 95)};
(2) \textbf{Specialization} means that a specific authoring task is always supported by the same modality. For example, Lyra2~\cite{zong_lyra_2021} uses the visual builder for visual mapping and VbD for view transformations (interactions);
Complementarity lets users employ multiple modalities which are merged together to perform one task. We make a temporal distinction between simultaneous and sequential combinations: (3) \textbf{complementary simultaneous} approaches, e.g., InChorus~\cite{srinivasan_inchorus_2020}, let users tap an attribute while pointing on the x-axis title region to bind the data attribute to the x-axis, while in (4) \textbf{complementary sequential} methods, e.g., in DataBreeze~\cite{srinivasan_interweaving_2021}, one select a set of points with touch, pause, and then issue a spoken command.

\section{Design Space}

\begin{figure*}%[!t]%
\centering
\includegraphics[width=1\linewidth, alt={}]{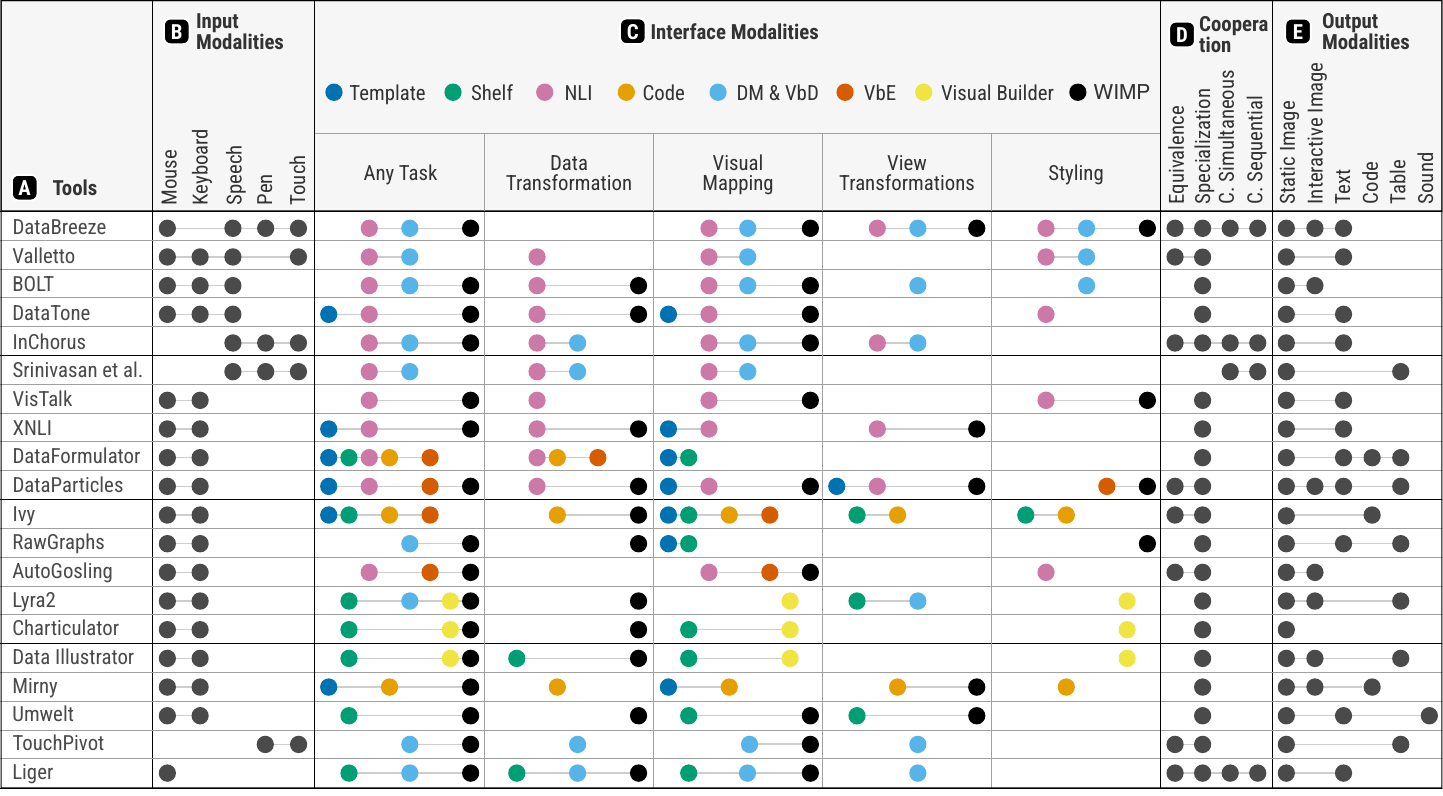}
\caption{Classification of multimodal interaction in \nrTools visualization authoring tools. Rows are the visualization authoring tools (A). Columns represent the five dimensions---Input Modalities (B), Interface Modalities and Authoring Tasks (C), Cooperation Methods (D) and Output Modalities (E)---and their categories. 
% \For each tool, we denote applicable categories per dimension with filled \circles}. Additionally, the Authoring Tasks dimension illustrates the supported interfaces and inputs per task. }
The rows are ordered based on the number of input modalities used in each tool (B).}\label{fig1}
\end{figure*}

To construct the design space, we adopted existing classifications (see Section~\ref{sec:classifications}) and additionally coded input and output modalities of the collected papers to describe the capabilities of the tools. This approach has resulted in a space of five distinct dimensions---input modalities, interface modalities, authoring tasks, cooperation methods and output modalities---to characterize multimodal interaction in visualization authoring tools~(\cref{fig1}). We discuss observated patterns and ambiguities for each dimension below. The survey results are also accessible via \url{https://gosling-lang.github.io/multimodal-vis-authoring}.

\subsection{Input Modalities~(Fig.~\ref{fig1}B)}
We observed five categories of input modalities for visualization authoring tools: mouse, keyboard, speech, pen, and touch. This observation aligns with the reported affordances of inputs for creating visualizations by Badam et al.~\cite{badam_affordances_nodate}. Not surprisingly, the majority (16/20) of tools support mouse and keyboard input since authoring tools are mainly developed for PC devices. Furthermore, we observe systems that provide speech and touch (4/20), and some additionally include pen input (3/20). 
% The (speech, touch, pen) tools are tablet-based or for large vertical screens, therefore typically not supporting mouse and keyboard. 
The (speech, touch, pen) tools are typically designed for tablets or large vertical screens and therefore do not support mouse and keyboard input.
Examples of the latter group are InChorus~\cite{srinivasan_inchorus_2020} and DataBreeze~\cite{srinivasan_interweaving_2021}. Two tools support four input modalities~\cite{kassel_valletto_2018, srinivasan_interweaving_2021}, and no tool supports all five. 

%Touch: This refers to the common issue in post-WIMP interfaces, since there is no graphical indicator for a function (e.g., button).

\subsection{Interface Modalities~(Fig.~\ref{fig1}C)}
The number and combinations of interface modalities vary widely across tools. Leaving general WIMP-like interface interactions out, which are typically needed to handle further refinements or resolve ambiguities in the expressed intents, most tools (13/20) support two interfaces. Natural Language Interfaces (NLI) are combined with direct manipulation (DM) or visualization-by-demonstration (VbD) interfaces in 5/20 tools. In all five, natural language is supported by speech input. For example, Valletto~\cite{kassel_valletto_2018} aims to support data-specific intents through speech and visual encoding intents through multitouch gestures. 
We identified an emergent interface modality, \textbf{visualization-by-example} (\textbf{VbE}), which allows users to compose or submit an example (e.g., an image or textual specification) for further use in the authoring proces, i.e., the system generates a reusable visualization template from the example input. We observed VbE interfaces (3/20) mainly in tools that also support template editing and often include NLI, e.g., DataParticles~\cite{cao_dataparticles_2023}. Intuitively, each interface modality may offer different affordances depending on the task.  We detail these task--interface patterns below. 

% \begin{itemize}
%     \item \blue{RawGraphs and DataFormulator have a special type of mix between template and shelf}
%     \item \blue{What are fan interfaces, interactive query widgets, concept shelves, block-based editors?}
%     \item recommendations shown or not? E.g., Mirny and Liger show ..., but other systems that use an AI agent do not always show alternatives 
% \end{itemize}

\subsection{Authoring Tasks~(Fig.~\ref{fig1}C)}

% \begin{itemize}
%    \item \blue{Add fourth task: styling}
   % \item \blue{Filtering being data transformation or a view transformation} --> discussion
%\end{itemize}

\paragraph{Data Transformation} Many tools allow users to perform data transformation tasks with NLI (9/20). In most cases, the NLIs---allowing input through speech, text, or both---tend to be complemented by a VbD/DM or WIMP strategy. The least frequently supported are shelf (2/20) and the new VbE (1/20) interfaces. No tool uses templates for data transformation. A notable tool specifically addressing data transformation burdens is Data Formulator~\cite{wang_data_2023}. It combines three interfaces (NLI, code editing, and VbE) as a new paradigm: concept binding. 

\paragraph{Visual Mapping} Similar to data transformation, NLIs are also commonly used for visual mapping (10/20 tools). Especially for this task, they tend to be combined with WIMP (7/20) or DM/VbD (5/20). Furthermore, we observe that template editors (7/20) and shelf construction interfaces (7/20) are employed for visual mapping tasks. Interestingly, we identified two tools~\cite{wang_data_2023, mauri_rawgraphs_2017} that use a mix between template and shelf configuration, with considerable freedom to map various channels but always using a top-down approach, directed by the visualization type. Data Formulator~\cite{wang_data_2023} focuses on data transformation and, therefore, does not expose the most expressive encoding options in its interface. RAWGraphs~\cite{mauri_rawgraphs_2017}, however, is aimed at creating visual mappings that are designed to be further modified in other tools. Another noteworthy tool is Ivy~\cite{mcnutt_integrated_2021}, which lets users create visual mappings through four modalities, including VbE by copying a Vega-Lite~\cite{satyanarayan_vega-lite_nodate} specification as a starting point. 

\paragraph{View Transformations}
We found the least support by all interface modalities for this task. In cases where there is, systems mostly use DM (5/20), sometimes with NLI or shelf configuration interfaces. Lyra 2~\cite{zong_lyra_2021} supports creation of brushing and linking interactions through a combination of shelf and DM interface modalities. DataParticles~\cite{cao_dataparticles_2023} and DataBreeze~\cite{srinivasan_interweaving_2021} each show three interaction modalities for this task. DataParticles is a one of a kind tool in this category, facilitating users in authoring data stories with complex animations through template, NLI and WIMP interface modalities.

\paragraph{Styling} In surveying the tools, we added styling as a separate task as it is often part of  users' customization goals. We define styling as operations on visual structures to refine aesthetics or emphasis without altering the underlying visual mapping. An example styling task is to add a border around points in a scatterplot. We observed that interfaces for styling tend to be a combination of NLI and DM (3/20) because of ease of use in the overall authoring flow. Furthermore, visual builder interfaces (3/20) offer great expressiveness to handle many styling cases, therefore typically sufficing as unimodal interfaces for this task.

\subsection{Cooperation Methods~(Fig.~\ref{fig1}D)}

In coding the cooperation methods, we considered both combinations of interfaces and input modalities. Nearly all of the tools we surveyed employ a specialization cooperation method (19/20), with the only exception being the design in the study by Srinivasan et al.~\cite{srinivasan_facilitating_nodate}. A considerable number of tools also demonstrated equivalence multimodal interaction (7/20). For example, Liger~\cite{saket_liger_2019} allows VbD to map an attribute to color by coloring a set of sample points or by dragging and dropping the data attribute to the color shelf. We also identified complementary simultaneous and sequential cooperation modes (4/20), though to lesser extents. This might in part be because of the difficulty in distinguishing it from specialization due to unclear task specifications and demonstrations of the tools. Most tools that offer complementary simultaneous interaction also provide the sequential variant. For example, DataBreeze~\cite{srinivasan_interweaving_2021} allows pointing on a location while speaking (simultaneous) or point and lift their finger---think---and then use speech (sequential). 3/20 tools in our survey support 4 different ways of cooperation between modalities, likely to cater to varying interaction patterns among users. This aligns with the observation (myth \#8) by Oviatt et al.~\cite{oviatt_ten_1999} that not all users integrate multimodal commands uniformly; some combine inputs simultaneously, while others prefer sequential cooperation.
Finally, even though not implemented in any tool, some tools note on the possible value of redundancy interactions~\cite{martin_tycoon_nodate} in order to prevent usability problems and support the learning process~\cite{kassel_valletto_2018}.

\subsection{Output Modalities~(Fig.~\ref{fig1}E)}
We also investigated the outputs of tools and identified differences in the representations exposed to the user. Aside from static visual output, many tools also provide text output (11/20), and some offer interactive visual output (7/20). Text usually represents some feedback message on the authoring process; for example, Liger~\cite{saket_liger_2019} explains each recommendation made in the VbD in natural language.

In a few cases, we observed the output of underlying elements that constitute the visual representation. Some tools also expose data (7/20) and code (3/20) as alternative representations that can be considered complementary modalities to help the user during the authoring process. The tool Umwelt~\cite{zong_umwelt_2024} provides sound output in addition to images and text for improved accessibility.

% \subsection{Patterns across Dimensions}

\section{Discussion}
% Here we discuss design implications and limitations in our approach.

%discussion

%limitations
The characterization of multimodal interaction was developed through systematic review of literature and iteratively refined based on frequent discussion among the authors. Although most design dimensions were selected to demonstrate distinct patterns, many tools exhibit capabilities that span multiple categories due to  the required versatility of these tools. Furthermore, we observed several ambiguities that may have lead to additional overlapping classifications. One notable example concerns the distinction between the cooperation strategies specialization and sequential cooperation when applied to authoring. If authoring is considered a single task, all tools in this survey would fall under \textit{sequential cooperation}. However, considering the high level tasks~(\cref{sec:classifications}), we could minimize the overlap. In cases of ambiguity, we relied on the author's claims to avoid over interpretation that would be challenging to ground in the paper. 

In surveying the interfaces for data transformations, we noticed that distinguishing between tools and their approaches for data and view transformations was not always trivial. Following the definition by Card et al.~\cite{card_1999}, data filtering is a form of view transformation. In the context of visualization authoring tools, we categorize data filtering that uses DM on visual structures as a view transformation. In contrast,  interfaces that do not directly interact with the visual encodings, such as using mouse input on WIMP (GUI) widgets~\cite{mcnutt_integrated_2021}, will be data transformations.

% limited study limitation, reliance on author claim
% futher investigation into which cooperation methods and which interfaces help the most for which tasks or for which users
% effect of the type of visualization being contstucted
% interactive visualizations with different types of interfaces, but most oten mouse, keyboard. opportunities for alterntative inputs for interaction?
% beyond visulization, how it is represented 

% \begin{itemize}[noitemsep]
%     \item \blue{bias in paper selection}
%     \item \blue{limitations of criteria?}
% \end{itemize}

% \subsection{\orange{some subtitle}}
% \blue{
% \begin{itemize}[noitemsep]
%         \item What is multimodality for visualization authoring? Unclear definition
%         \item 
% “Some modalities may not have obvious points of similarity and straightforward ways to connect.”
% \item Multimodality may reduce or increase the cognitive load.
% Discoverability
% Learnability 
%     \item What are the pro’s and con’s of different authoring outputs -- mutlimodal output or representations? Data and code help to understand
% \item beyond the output that is exposed/ shown to the user, how do we represent visualizations? What implications does it have for building visualizations? Search engines can find similar visualizations? 
% \end{itemize}
% }
\section{Conclusion}
We have proposed a design space to characterize multimodal interactions in visualization authoring tools. We reviewed \nrTools tools and identified five dimensions which describe multimodal interactions according to the inputs, interfaces, outputs, cooperation strategies, and tasks they aim to support. We found a large variety of tools in these dimensions, and uncovered some ambiguities in defining aspects of multimodal interactions and visualization authoring that could benefit from additional research. We hope that this survey will serve as a valuable resource that future work can build upon to further characterize and expand our understanding of multimodal interactions in visualization authoring.

% \begin{itemize}[noitemsep]
%     \item \blue{main insight found}
%     \item \blue{take away message} 
%     \item \blue{future work}
% \end{itemize}

\acknowledgments{
This work was in part supported
by the National Institutes of Health (R01HG011773).}

\bibliographystyle{abbrv}

\bibliography{multimodal-vis-authoring}
\end{document}